\documentclass[prl,aps,twocolumn,showpacs,preprintnumbers]{revtex4-1}
\usepackage{latexsym,amsfonts,amssymb,amsmath,graphicx,graphics,hyperref}
\usepackage{amsfonts,amssymb,amsmath}            
\usepackage[]{graphics,graphicx,epsfig}            
\usepackage{amsthm}
\usepackage{epstopdf}

\def\identity{\leavevmode\hbox{\small1\kern-3.8pt\normalsize1}}
\newtheorem{theorem}{Theorem}

\newcommand{\ket}[1]{\left | #1 \right\rangle}
\newcommand{\bra}[1]{\left \langle #1 \right |}

\newcommand{\diag}{\text{diag}}
\renewcommand{\epsilon}{\varepsilon}
\newcommand{\1}{1\!\!1}

\newcommand{\be}{\begin{equation}}
\newcommand{\ee}{\end{equation}}
\newcommand{\bq}{\begin{eqnarray}}
\newcommand{\eq}{\end{eqnarray}}

\begin{document}

\title{Energy gaps of Hamiltonians from graph Laplacians}
\date{\today}

\author{Abbas \surname{Al-Shimary} and Jiannis K. \surname{Pachos}}
\affiliation{
School of Physics and Astronomy, University of Leeds, Leeds LS2 9JT,
U.K.}

\begin{abstract}

The Cheeger inequalities give an upper and lower bound on the spectral gap of discrete Laplacians defined on a graph in terms of the geometric characteristics of the graph. We generalise this approach and we employ it to determine if a given discrete Hamiltonian with non-positive elements is gapped or not in the thermodynamic limit. First, we define the graph that corresponds to such a generic Hamiltonian. Then we present a suitable generalisation of the Cheeger inequalities that overcomes scaling deficiencies of the original version. By employing simple examples we illustrate how the generalised Cheeger inequalities can successfully identify gapped or gapless phases and we comment on the computational complexity of this approach.

\end{abstract}

\pacs{02.10.Ox, 03.65.-w}

\maketitle

{\bf Introduction:} In physics, the variational approach plays a central role in theoretical and numerical approximations of spectral properties of Hamiltonians. When applied to the first excited state it can give an upper bound to the spectral gap of the Hamiltonian, provided that the ground state is known. Unfortunately, a lower bound to this gap is not readily available. Such a bound could help in understanding fundamental properties of many body systems that depend on the behaviour of the spectral gap in the thermodynamic limit \cite{Nielsen, Wim, Freedman}. A direct approach to this question is, in general, a hard computational problem as it requires the diagonalisation of exponentially large matrices. It is, thus, fascinating that the spectral gap of discrete Laplacians defined on a graph can be upper and lower bounded by general geometric characteristics of the graph \cite{Cheeger,Chung}. This is achieved with the help of the Cheeger constant that determines the extent a bottleneck configuration appears in a graph.

Here we derive a new version of the Cheeger inequalities that can estimate the energy gap of stoquastic Hamiltonian, i.e. Hamiltonians with all off-diagonal elements real and non-positive. For that we need to first associate a graph Laplacian to the Hamiltonian provided we know its ground state. This is not a major drawback as there are large families of physically relevant states, e.g. the Matrix Product States, that are ground states of Hamiltonians which are not known to be gapped or not in two or higher dimensions \cite{Schuch}. Such an important example is the two-dimensional AKLT model that can support universal quantum computation by measurements only, but it is not proven yet if it is gapped which would establish its fault-tolerance.

The upper Cheeger bound is based on the familiar to physicists variational method. To generalise the lower bound to the case of Hamiltonians we are forced to deviate from the usual methodology \cite{Chung} and employ the duality between the maximum flow and the minimum cut of a graph \cite{Alon}. This provides much greater flexibility and allows us to efficiently lower bound the spectral gap of a Hamiltonian in the thermodynamic limit. We also employ an additional structure, the reduced graph, that allows to optimise the lower bound beyond what is currently possible \cite{Chung}.

The methodology presented here sets the framework to study the spectral gap of Hamiltonians in terms of geometric characteristics of graphs without any approximations. Hence, the difficulty in estimating the energy gap of a Hamiltonian, when its ground state is known, has been translated to finding the Cheeger constant of the corresponding graph, which is known to be an NP-complete problem \cite{Golovach}. Our approach aims to give a new and general methodology that can facilitate the evaluation of energy gaps theoretically or numerically \cite{Freedman}.

{\bf Hamiltonians and Graph Laplacians:} We shall consider Hamiltonians, $H$, that are Hermitian $N\times N$ matrices with real non-positive entries. By employing the Perron-Frobenious Theorem ~\cite{Horn} it can be shown that the smallest eigenvalue $\lambda_0$ of $H$ is negative and all other eigenvalues are strictly larger in absolute value. Hence, we can denote the $N$ eigenvalues as 
$$
\lambda_0<\lambda_1\leq \lambda_2...\leq \lambda_{N-1}.
$$
Moreover, the eigenvector $\psi_0=(\alpha_i)_{i=0}^{N-1}$ corresponding to $\lambda_0$ is unique and all its components, $\alpha_i$, are positive satisfying $\sum_i|\alpha_i|^2=1$. We are interested in estimating the spectral gap of $H$ given by the difference between the lowest and second lowest eigenvalues 
\be
\Delta(H)=\lambda_1-\lambda_0.
\label{Gap}
\ee
The first excited state $\psi_1=(\beta_i)_{i=0}^{N-1}$ has components that are both positive and negative in order for the orthogonality condition, $({\psi_0},{\psi_1})=0$, to hold.

From the Hamiltonian, $H$, we next define the Laplacian operator $L$
\be
L = -\lambda_0\1_N+D^{-1}HD
=\begin{cases}
-\lambda_0 +H_{ii}, & i = j;\\
{\alpha_j\over \alpha_i}H_{ij}, & i\neq j.
\end{cases}
\label{Laplacian}
\ee
where $D$ is the diagonal matrix $D=\diag(\alpha_0,...,\alpha_{N-1})$ and $\1_N$ is the $N$-dimensional identity. In contrast to the usual paradigm \cite{Chung} Laplacian $L$ is a non-symmetric matrix that has a number of useful properties. Its rows sum up to zero [$\sum_jL_{ij}= -\lambda_0+\sum_j H_{ij}\alpha_j/\alpha_i=-\lambda_0+\lambda_0\alpha_i/\alpha_i=0$] and its lowest eigenvalue is $0$ with left eigenvector $\pi\equiv (\alpha_0^2,...,\alpha_{N-1}^2)$, i,e, $\pi L=0$, and right eigenvector $D^{-1}\psi_0=(1,...,1)^T\equiv\mathbf{1}$, i.e. $L\mathbf{1}=0$. Most importantly $L$ has the same energy gap as $H$, i.e. $\Delta(L)=\lambda_1-\lambda_0=\Delta(H)$ [$L(D^{-1}\ket{\psi_0})=0$ and $L(D^{-1}\ket{\psi_1})=(\lambda_1-\lambda_0)(D^{-1}\ket{\psi_1})$].

We now show how to construct a graph from the Laplacian $L$ ~\cite{Wim, Sinclair}. Consider a set of vertices $V=\{0,...,N-1\}$. For each pair of vertices $i,j$, which do not need to be distinct, the edge $(i,j)$ has a positive weight 
\be
w_{ij}=-\alpha_iH_{ij}\alpha_j.
\label{Weights}
\ee
Hence, the set of edges $E$ of the graph involve all $(i,j)$ for which $H_{ij}\neq 0$. As we have determined the vertices and the edges of the graph we can now define the degree of each vertex in the following way
\be
d_i =\sum_{j \in V} w_{ij} = -\alpha_i \sum_{j\in V}H_{ij} \alpha_j =|\lambda_0| \alpha_{i}^2.
\label{d}
\ee
Thus, we could think of $|\lambda_0|$ as the `bare' degree of $i$ and $d_i$ as the `dressed' degree of $i$. Similarly,  $-H_{ij}$ could be thought of as `bare' weight of the edge $(i,j)$ and $w_{ij}$ as the `dressed' weight of $(i,j)$. Note that $G$ is always connected because $H$ is irreducible. This completes our first task in identifying the appropriate graph and Laplacian for a given Hamiltonian, $H$. Next, we show how weighted graph can be used to give a lower bound to the spectral gap of the Hamiltonian. 

\begin{figure}[t]
\includegraphics[width=5cm]{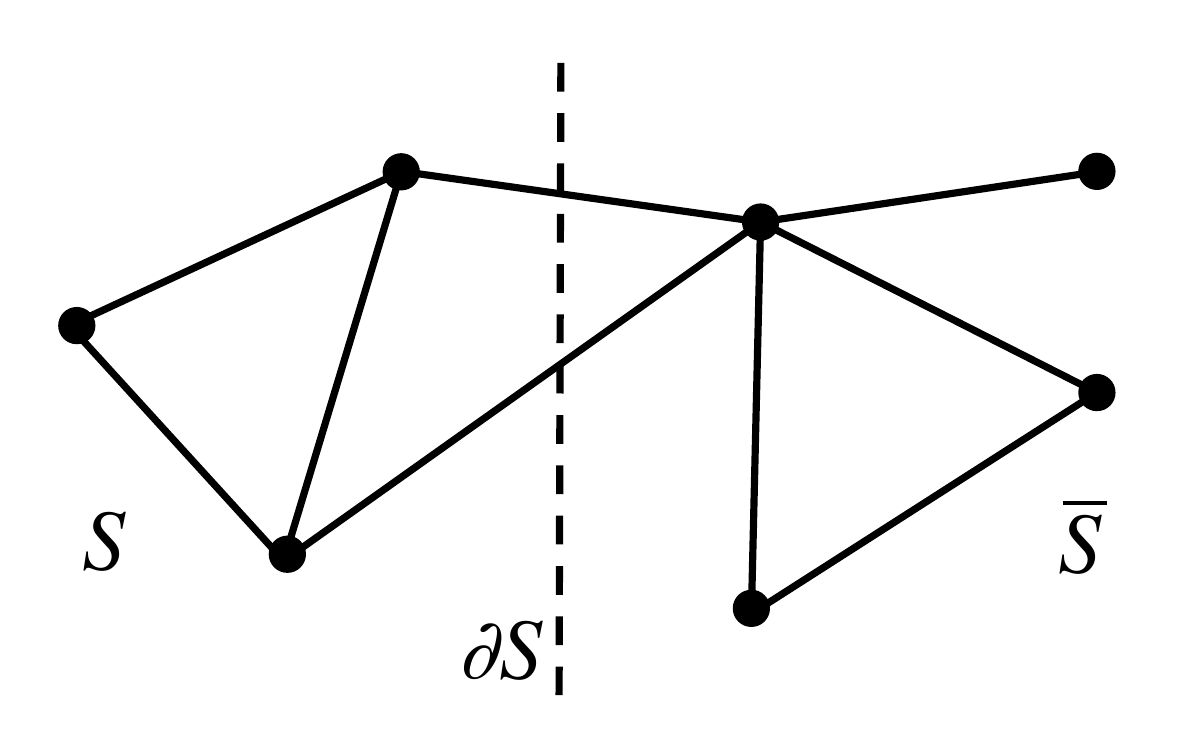}
\caption{An unweighted graph $G$ with vertices $V$ and edges $E$. The Cheeger cut, $\partial S$, is depicted that splits the vertices $V$ into $S$ and its compliment $\bar S$.}
 \label{fig:Graph}
\end{figure}

{\bf Cheeger constant and Cheeger inequalities:} We now introduce the appropriate Cheeger constant associated with a graph $G$. Consider a bipartition $S$ and $\bar S$ in the vertex set $V$ of the graph. Let the flow and the capacity be defined as
\be
F_S=\sum_{i \in S, j \in \bar{S}} w_{ij}\,\,\mathrm{and}\,\, C_S=\sum_{i \in S} \alpha_i^2,
\ee
respectively. We define the Cheeger constant as
\be
\Phi = \min_{\substack{S \\ C_S \leq 1/2}} {F_S \over C_S},
\ee
where the minimisation is over all possible partitions. The edge boundary $\partial S$ that corresponds to $\Phi$ is called here the Cheeger cut and it identifies the bottleneck of the graph, as depicted in Fig. \ref{fig:Graph}.

Next we present the Cheeger inequalities~\cite{Chung}. Consider the Laplacian operator (\ref{Laplacian}) with spectral gap $\Delta(L) =\lambda_1-\lambda_0$, defined on the graph $G$ with Cheeger constant $\Phi$. Then the Cheeger inequalities are give by
\be
\mathrm{(a)}\,\,2\Phi \geq \lambda_1 - \lambda_0\,\,\mathrm{and}\,\, \mathrm{(b)}\,\,
\lambda_1 - \lambda_0 \geq \frac{\Phi^2}{2|\lambda_0|}.
\label{Ineqs}
\ee
The Cheeger inequalities bound from above and below the spectral gap of the Laplacian and of the corresponding Hamiltonian. It is worth noting that the bounds are functions of the Cheeger constant $\Phi$ and the `bare' degree $|\lambda_0|$ which are geometric characteristic of the graph. Hence, if we had the means to determine the general shape of the graph from general properties of the Hamiltonian then we could successfully estimate the energy gap $\Delta(H)$.

The proof of the upper bound (\ref{Ineqs}a) is rather simple and it is based on a variational argument for the first excited state of the Laplacian~\cite{Chung}. By taking a vector $\psi$ that is orthogonal to the ground state of the Laplacian we guarantee that the expectation value of $L$ with respect to $\psi$ will be larger or equal to the gap $\lambda_1-\lambda_0$. Such a vector can be defined on a bipartition of $G$ into $A$ and $B$ so that
\be
\psi_i =  
\begin{cases}
{1 \over C_A}, & i \in A;\\
-{1\over C_B}, & i\in B.
\end{cases}
\ee
The variational parameter is the position of the boundary that separates $A$ and $B$. The optimal value is obtained when the boundary is identical to the Cheeger cut, where the gap is upper bounded by $2\Phi$ as shown in (\ref{Ineqs}a). Hence, Hamiltonians that give rise to graphs with a predominant bottleneck behaviour, i.e. small $\Phi$, have a small gap. If on the other hand $\Phi$ is large it does not automatically guarantee that the Hamiltonian has a large gap. This is due to the factor $1/|\lambda_0|$ in the lower bound (\ref{Ineqs}b), where $|\lambda_0|$, roughly speaking, corresponds to the size of the system. As we are interested in the behaviour of the lower bound in the thermodynamic limit we are faced with the task of improving this bound.

{\bf Two characteristic examples:} We shall now see with explicit examples how the Cheeger inequalities are applied to physical systems. Our first example concerns a free particle hoping on a one dimensional lattice. For simplicity we take the lattice to have periodic boundary conditions and to be of size $N$. The Hamiltonian is given by 
\be
H=-t\sum_{i=1}^N \big(\ket{i}\bra{i+1}+\ket{i+1}\bra{i}\big),
\ee
where the state $\ket{i}$ denotes the particle being in position $i$ with the periodic condition $\ket{N+1}=\ket{1}$. The ground state is given by $\ket{\psi_0} = {1 \over \sqrt{N}}\sum_{i=1}^N \ket{i}$ with eigenvalue $\lambda_0 =-2t$. By Fourier transformation one can show that $\lambda_1 =-2t\cos(\pi/N)$. Hence, in the thermodynamic limit $\Delta(H)=2t\big[1-\cos(\pi/N)\big]\approx t\pi^2/N^2\rightarrow 0$ as $N\rightarrow \infty$.

\begin{figure}[t]
\includegraphics[width=5.5cm]{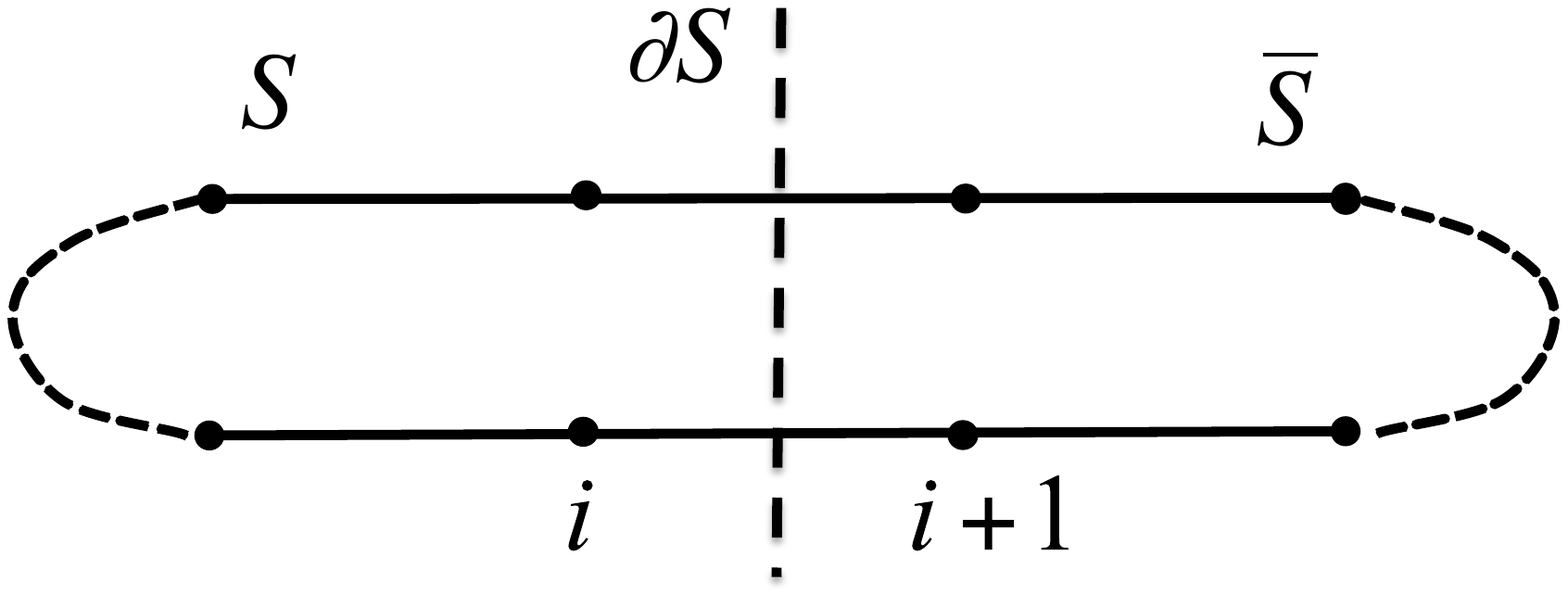}
\caption{The graph corresponding to a particle in one dimensional lattice of $N$ sites with hopping amplitude $t$ subject to periodic boundary conditions. For this system the graph as well as the physical lattice configuration are the same. Symmetry considerations show that the Cheeger constant is $\Phi=4t/N$.}
 \label{fig:line}
\end{figure}

Let us now turn to the graph theoretic approach of this system. Following definitions (\ref{Laplacian},\ref{Weights},\ref{d}) we obtain the graph of Fig. \ref{fig:line} with Cheeger constant $\Phi =4t/N$. The Cheeger inequalities give
\be
{8 t\over N} \geq \lambda_1-\lambda_0 \geq {4 t \over N^2},
\ee
that is in agreement with the exact result. Hence, the estimation of the energy gap from the Cheeger inequalities gives the gapless asymptotic behaviour of the one dimensional free particle when $N\rightarrow \infty$.

Next we consider the Hamiltonian of non-interacting spin-$1/2$ particles
\be
H = -B\sum_{i=1}^n \sigma^x_i,
\label{mag}
\ee
where $\sigma^x$ is the Pauli operator and $B>0$. The ground state is an equal superposition of all spin states $\ket{\psi_0}={1\over\sqrt{2^n}}\sum_{i=1}^{2^n}\ket{i}$ where $\ket{i}$ denotes a particular spin configuration among the total of $N=2^n$ in the $\sigma^z$ eigenbasis. The two lowest eigenvalues are $\lambda_0=-Bn$ and $\lambda_1=-B(n-2)$ giving the gap $\Delta(H)=2B$, which is constant as $n$ is taken to the thermodynamic limit. 

\begin{figure}[t]
\includegraphics[width=8.5cm]{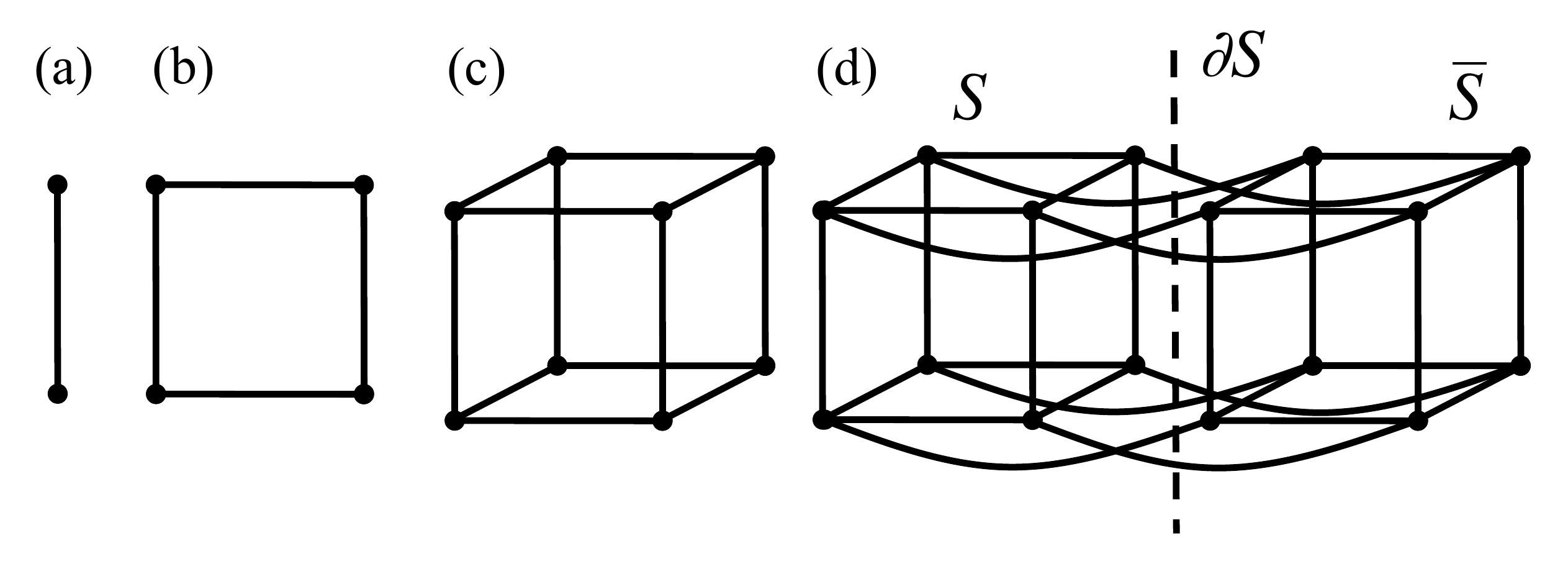}
\caption{The hypercube graphs $Q_n$ for (a) $n=1$, (b) $n=2$, (c) $n=3$ and (d) $n=4$. In all cases one can see that the optimal cut, $\partial S$, has $S$ and $\bar S$ being lower dimensional hypercubes $Q_{n-1}$.}
 \label{fig:square}
\end{figure}

The corresponding graph is a hypercube $Q_n$ depicted in Fig. \ref{fig:square}. The weights of the edges are all $B$ so that the Cheeger constant is $\Phi = B$. The Cheeger inequalities give
\be
2B\geq \Delta(H)\geq {B \over 2n}.
\ee
We see that, while both inequalities hold, the lower bound goes to zero when $n\rightarrow \infty$. Hence, the Cheeger inequality gives no information if the Hamiltonian $H$ is gapped or not in the thermodynamic limit. This simple example demonstrates that the current lower bound is not adequate for our purposes. In the following we introduce a generalisation of the lower bound of the Cheeger inequalities that can recognise if a Hamiltonian is indeed gapped or not in the thermodynamic limit. 

{\bf Generalised Cheeger inequality:} The inadequacy of the lower bound (\ref{Ineqs}b) comes from the factor $1/|\lambda_0|$, where $|\lambda_0|$ corresponds to the maximum number of edges originating from a vertex. In the following we show that it is not necessary to consider all edges of the graph $G$ in the derivation of the lower bound. In fact, a generalised form of the lower bound can be derived that does not suffer from the scaling problem of the original version and, hence, it can be applied to problems of physical interest. 

Let us assume we know the optimal bipartition of a graph $G$ in $S$ and $\bar S$ by the Cheeger cut $\partial S$. We consider a reduced version of the graph $G$, named $\tilde G$, that has exactly the same vertices as $G$, but a reduced set of edges, $\tilde E$. The `bare' degree of vertices in $\tilde G$ are given by $c_i = \sum_{j,(i,j)\in \tilde E} w_{ij}/\alpha_i^2$. Hence, $\tilde G$ has a smaller  maximum `bare' degree than $G$ namely, $c \equiv\max _{i\in V} c_i \leq |\lambda_0|$. Consider now all possible subsets $S_i$ of the set of vertices of $S$. For each subset we define the ratio $\tilde \Phi_i = \tilde F_{S_i}/C_{S_i}$ where $\tilde F_{S_i}$ is defined only in terms of the edges of the reduced graph and we set $\tilde \Phi=\min_{{S}} \tilde \Phi_i$. Note that, unlike $\Phi$, the minimisation in the definition of $\tilde \Phi$ is performed only over the set $S$ not over the whole $G$. Nevertheless, since the reduced graph has a smaller number of edges, the value of $\tilde \Phi$ might be actually smaller than $\Phi$.

With the help of the reduced graph one can derive (see Supplement) a generalised version of the lower bound:
\be
\lambda_1-\lambda_0 \geq {\tilde \Phi^2 \over 2c}.
\label{generalised}
\ee
This inequality shows that the energy gap of $H$ can be bounded by geometric characteristics of a reduced graph $\tilde G$. Initially, note that for $\tilde G\equiv G$ the lower bounds (\ref{Ineqs}b) and (\ref{generalised}) are identical. As the reduced graph can be arbitrarily chosen we aim to define $\tilde G$ in order to maximise the lower bound (\ref{generalised}). If a reduced graph can be found with a lower Cheeger bound that does not tend to zero in the thermodynamic limit then the Hamiltonian is gapped. 

\begin{figure}[t]
\includegraphics[width=8.5cm]{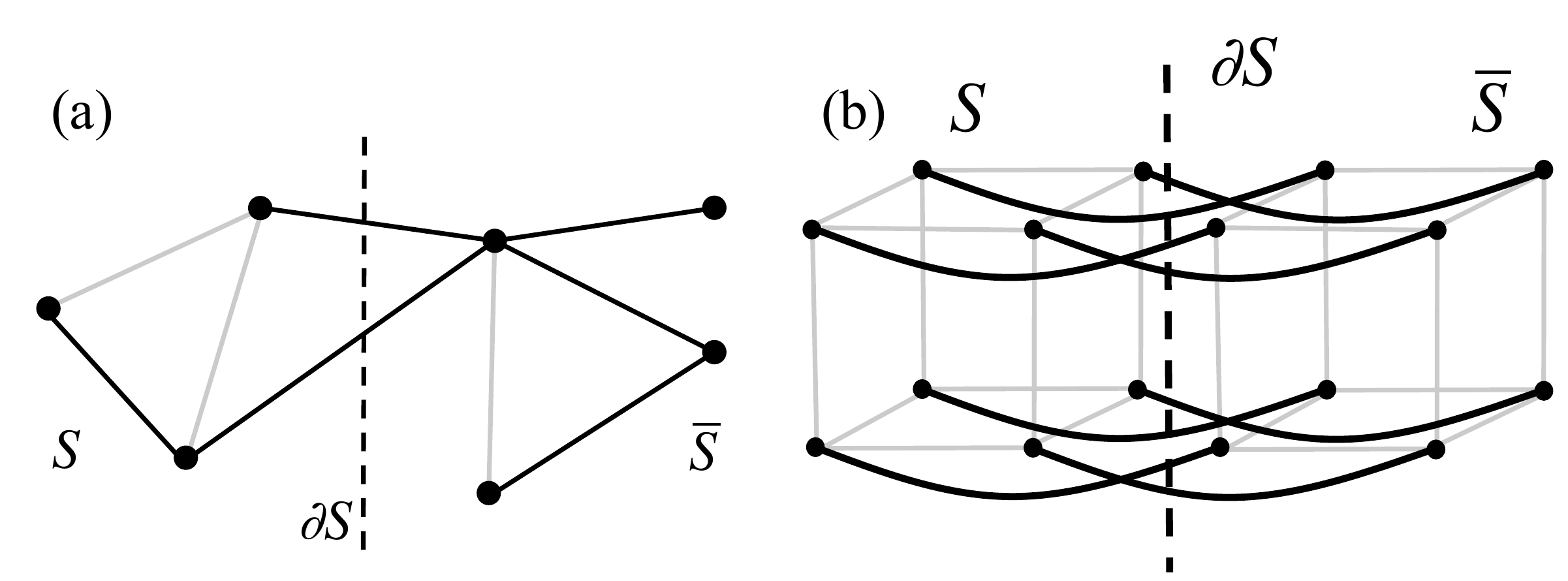}
\caption{Possible reduced graphs for (a) the general graph of Fig. \ref{fig:Graph} and (b) for the $Q_4$ hypercube. These reduced graphs are chosen to optimise the lower Cheeger bound. The removed edges are denoted in light grey.}
 \label{fig:squareExample}
\end{figure}

The advantage of this approach is that for certain Hamiltonians we can easily make $c$ finite as the size of the system, $N$, increases. This can be performed by having only a fixed `bare' degree in $\tilde G$ that does not increase with $N$. Moreover, it is desired to have $|\partial S_i|\neq0$ for all subsets $S_i$ so that the inequality (\ref{generalised}) does not become trivial. To meet these requirements a general rule can be adopted of keeping in $\tilde E$ all edges that belong in the Cheeger cut and paths of edges that connect all vertices in $S$ with vertices that have edges in $\partial S$ (see for example Fig. \ref{fig:squareExample}(a)). 

Even if at a first sight the choice of the reduced graph might seem random it is rather straightforward to demonstrate its versatility. Consider, for example, the case of the simple spin-$1/2$  model given in (\ref{mag}). The energy spectrum of this model is gapped in the thermodynamic limit, but the usual Cheeger inequalities failed to demonstrate this. Let us define the reduced graph such that all edges are removed except the ones that belong to the Cheeger cut (see Fig. \ref{fig:squareExample}(b)). This has $c=B$ and $\tilde \Phi =\Phi=B$ giving the lower bound
\be
\lambda_1-\lambda_0\geq{B\over 2},
\ee
which is independent of the system size. Hence, the generalised Cheeger inequality can successfully prove the gapped nature of this model in the thermodynamic limit.

\begin{figure}[t]
\includegraphics[width=4.27cm]{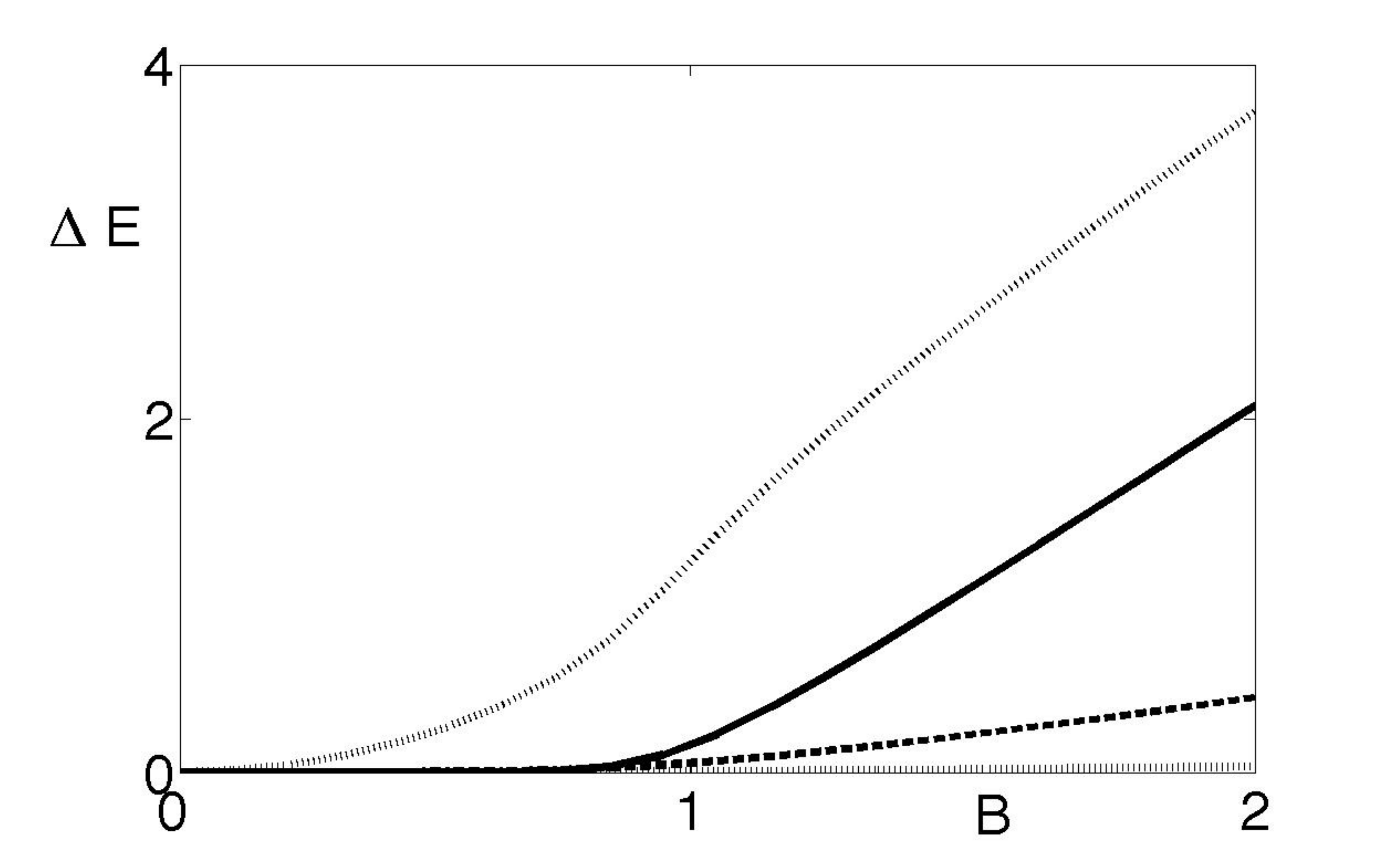}
\includegraphics[width=4.27cm]{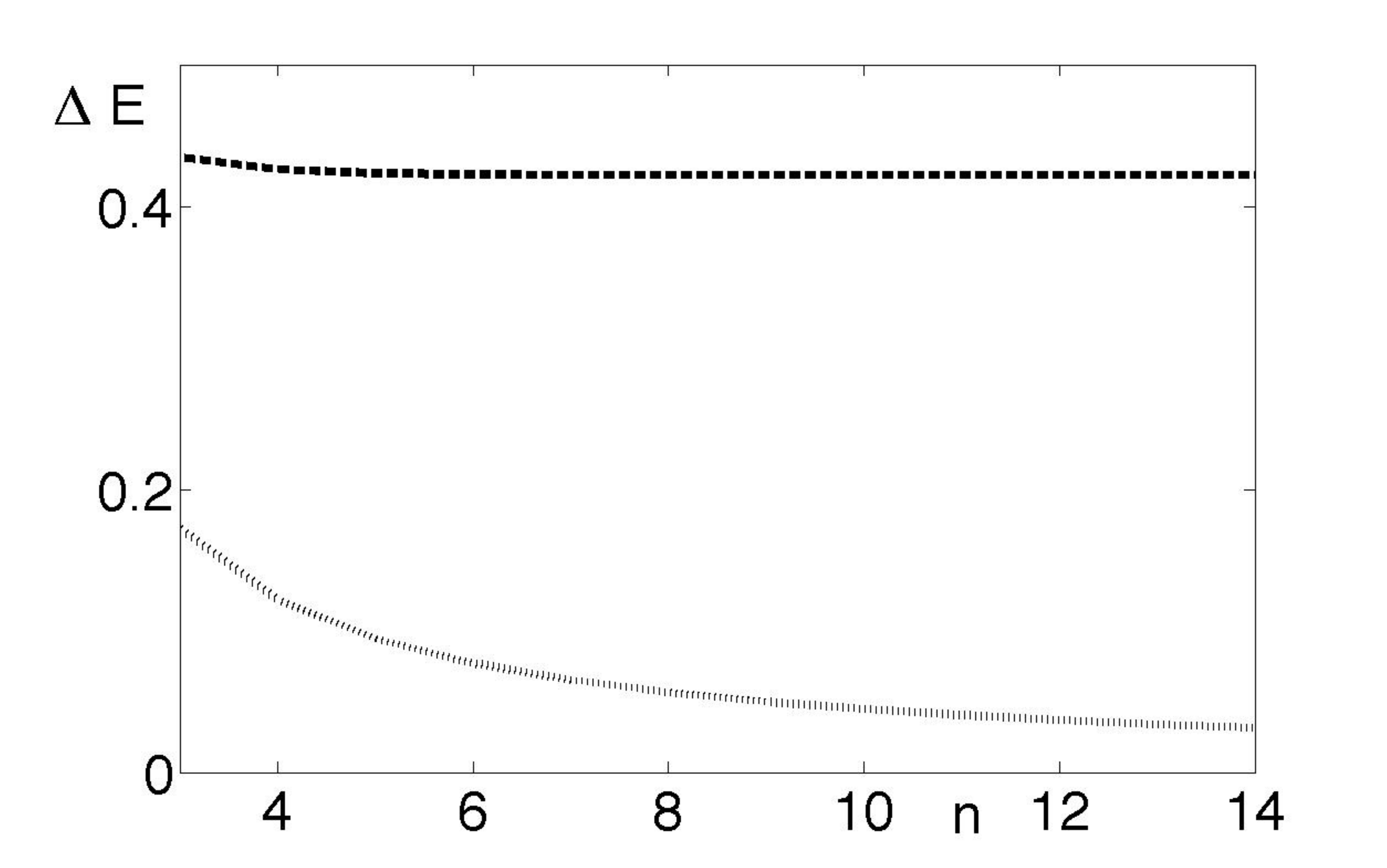}
\caption{The upper and lower bounds of the energy gap for the Ising model. (a) The spectral gap (solid line) for $n=14$. The upper and lower Cheeger bounds (dotted lines) and the generalised lower bound (dashed line) as a function of the magnetic field, $B$. (b) The lower Cheeger bound (dotted line) and the generalised lower bound (dashed line) as a function of the system size $n$ for $B=2$. It is clearly seen that, unlike the usual Cheeger bound, the generalised one successfully bounds the gap as the system size increases.}
 \label{fig:Ising}
\end{figure}

Consider now the one dimensional Ising model
\be
H= -\sum_{i=1}^{n-1} (\sigma_i^z\sigma_{i+1}^z+2\1_4)-B\sum_{i=1}^{n}\sigma^x_i.
\ee
As this is a spin-$1/2$ Hamiltonian it gives rise to a hypercube graph. At $B=0$ it has a doubly degenerate ground state, at $B=1$ it exhibits critical behaviour as $n\rightarrow \infty$ and it has a unique gapped ground state at $B>1$. Fig. \ref{fig:Ising}(a) illustrates the usual and generalised Cheeger bounds for various values of the magnetic field $B$. Fig. \ref{fig:Ising}(b) shows that the usual Cheeger bound scales unfavourably as a function of $n$, while the generalised one remains almost constant as $n$ increase, successfully bounding the energy gap from below. This general prescription can be straightforwardly applied to determine theoretically or numerically the gapped nature of more complex systems~\cite{Abbas}.

{\bf Conclusions:} In this Letter we generalised the variational method that determines an upper bound for the energy gaps of Hamiltonians to the case of lower bounds. To upper bound the energy gap one employs the variational ansatz for the first excited state. This method introduces variational parameters as degrees of freedom with which one can optimise the bound. Similarly, in the case of the lower bound choosing the reduced graph $\tilde G$ provides the freedom for optimisation. In the presence of ground state degeneracy due to a symmetry one can project the Hamiltonian to irreducible components and treat separately each of them. We envision that the presented method can actually reveal the behaviour of the gap in the thermodynamic limit of a wide variety of physical problems ranging from condensed matter to high energy physics.

{\bf Acknowledgements:} We would like to thank Wim van Dam,  Michael Freedman, Jonathan Keating, Nicholas Read and Kirill Shtengel for inspiring conversations. This work was supported by EPSRC and the Royal Society.

\section{Supplement}

Recall that for a Hamiltonian $H$, the corresponding Laplacian is given by
\[L = -\lambda_0\1_N+D^{-1}HD.\]
Let $\psi_1$ be an eigenvector of $H$ with eigenvalue $\lambda_1$, so $e=\psi_1^{T}D=(e_i)^{N-1}_{i=0}$ is a left eigenvector of $L$ with eigenvalue $\lambda_1 <\lambda_0$.  Then
\begin{equation} 
e L=(\lambda_1-\lambda_0)e.
\label{inner}
\end{equation}
Define the subset of vertices 
\[ V^{+}=\{ i \in V: e_i>0\}.\] 
Note that $\sum_{i\in V}e_i=0$ because
\[0=(\psi_1,\psi_0) = (\psi_{1}D, D^{-1} {\psi_{0}})=(e,\mathbf{1})=\sum_{i\in V} e_{i}.\]
We may assume without loss of generality that $C_{V^+}=\sum_{i\in S} \alpha^2_i\leq \frac{1}{2}$. Now let $\hat{e}$ be the vector defined by 
\[\hat{e}_i=
\begin{cases}
e_i/\pi_i, & i \in V^{+}; \\
0, & \text{otherwise.}
\end{cases}
\]
We assume that $\hat{e}_0 \geq \hat{e}_1\geq \ldots \hat{e}_{N-1}$, which implies that $V^+=\{0,1,\ldots,r \}$ for some $0 \leq r <N-1$. Taking the scalar product of \eqref{inner} with $\hat{e}$ gives
\begin{equation}
(eL,\hat{e})=(\lambda_1-\lambda_0)(e,\hat{e})=(\lambda_1-\lambda_0)\sum_{i \in V^+} \pi_i \hat{e}^2_i.
\label{rhs}
\end{equation}
The left hand side can be expanded as
\be
(eL,\hat{e}) = \sum_{i\in V^+} \sum_{j\in V} \hat{e}_i L_{ji} e_j
\geq \sum_{i \in V^+} \sum_{j \in {V^+}}\hat{e}_i L_{ji} e_j, 
\ee
since both $L_{ji}$ and $e_j$ are negative for $j \notin V^+$;
\be
...=\sum_{i \in V^+} \sum_{\substack{j \in V^+ \\j\neq i}} \hat{e}_i ({\alpha_i\over \alpha_j}H_{ji}) {e_j} - \sum_{i \in V^+} \sum_{j \neq i} \hat{e}_i ({\alpha_j\over \alpha_i} H_{ij}) {e_i}, 
\ee
by $L_{ji}={\alpha_i\over \alpha_j}H_{ji}$ and $L_{ii}=-\lambda_0+H_{ii}=-\sum_{j \neq i} {\alpha_j\over \alpha_i}H_{ij}$;
\be
...=-\sum_{i \in V^+} \sum_{\substack{j \in V^+ \\ j \neq i}} w_{ij} \hat{e}_i \hat{e}_j + \sum_{i \in V^+} \sum_{j \neq i} w_{ij} \hat{e}_i^2, 
\ee
by $w_{ij}=-\alpha_i H_{ij} \alpha_j$;
\be
...=-2 \sum_{i<j} w_{ij} \hat{e}_i \hat{e}_j + \sum_{i<j} w_{ij} (\hat{e}_i^2 + \hat{e}_j^2), 
\ee
by using $\sum_{\substack{i,j\\ i\neq j}}=2\sum_{i<j}$;
\be
...=\sum_{i<j} w_{ij} (\hat{e}_i-\hat{e}_j)^2.
\label{lhs}
\ee
Putting together \eqref{rhs} and \eqref{lhs} we get
\begin{equation}
\Delta(L)=\lambda_1-\lambda_0 \geq \frac{\sum_{i<j} w_{ij}(\hat{e}_i-\hat{e}_j)^2}{\sum_{i\in V^+} \pi_i \hat{e}^2_i}.
\label{lowerbound}
\end{equation}

Now we are ready to prove the following theorem.
\begin{theorem} For the the Laplacian $L$ given in (\ref{Laplacian}) it holds that
\[  \lambda_1 - \lambda_0 \geq \frac{\tilde{\Phi}^2}{2 c},\]
where the quantity $\tilde{\Phi}$ is defined so that the reduced flow out of $S$, $\tilde{F}_{S}$ satisfies 
\[ {\tilde{F}_{S}} \geq \tilde{\Phi}C_{S} \] 
for any $S\subset V$ and,
\begin{align}
\tilde{F}_{S} &=\sum_{\substack{i\in S, j \in V \\ (i,j) \in \tilde E }} w_{ij}, \; \text{the reduced flow out of $S$;}\nonumber\\
c&=\max_{i \in V} ({\sum_{j,(i,j)\in \tilde E} w_{ij}} /\pi_i), \; \text{the constriction.}\nonumber
\end{align}
\end{theorem}

\textbf{Proof}: To prove Theorem 1 we employ the max-flow min-cut theorem \cite{Alon}. Unlike the standard proof that is based on a specific vertex enumeration \cite{Chung}, the max-flow min-cut theorem provides a much more versatile framework that allows us to successfully generalise the lower bound. Consider the network $N$, based on the graph $G$, with vertex set $ \{s,t \} \cup X \cup Y$ where $s$ is the source, $t$ is the sink, $X$ is a copy of $V^+$ and $Y$ is a copy of $V$. The directed edges of this network and their capacities are given as follows:
\begin{enumerate}
\item For every $x \in X$, the directed edge $(s,x)$ has capacity $(1+\tilde{\Phi}) \pi_i$ where $x$ is labelled by vertex $i$.
\item For every $x \in X$, $y \in Y$, there is a directed edge $(x,y)$ with capacity $w_{ij}$ if $x$ is labelled by a vertex $i$,  $y$ is labelled by a vertex $j$ and $(i,j)$ is an edge in $G$ such that $(i,j)\in \tilde E$. Otherwise we set the capacity of $(i,j)$ to be zero.
\item For every $x \in X$, $y \in Y$, there is a directed edge $(x,y)$ with capacity $\pi_i+w_{ii}$ if $x$ and $y$ are labelled by the same vertex.
\item For every $y \in Y$ labelled by $j$, the directed edge $(y,t)$ has capacity $\pi_j$.
\end{enumerate}

We claim that the value  of the min-cut of this network is $(1+\tilde{\Phi}) C_{V^+}$, where $C_{V^+}=\sum_{i \in V^+} \pi_i$. To show this, let $K$ denote a cut separating $s$ and $t$. Let 
\begin{align}
X_{1} :&= \{ x \in X : (s,x ) \notin K \}\nonumber\\
Y_1 :&=\{ y \in Y: (y,t) \in K \}.
\nonumber
\end{align}
Define the following sets of edges,
\begin{align}
E_X&:= \{ (s,x) : x \in X \},\nonumber\\
E_{X_1} &:= \{( s,x ): x \in X_1 \},\nonumber\\
E(X_1,X_1) &:= \{ ( x,y ): x \in X_{1} \subset X  \, \text{and} \, y \in  X_{1} \subset Y   \},
\nonumber
\end{align}
The total capacity of the cut $K$ is at least the sum of capacities of the edges,
\begin{itemize}
\item Edges $E_X \setminus E_{X_1}$,
\item The edges $E(X_1,X_1)$,
\item The edges $E(X_1,Y)$.
\end{itemize}
Therefore, the total capacity of the cut is at least equal to
\begin{align}
&\,\,\,\,\,\,\,\,(1+ \tilde{\Phi}) C_{X - X_1} + C_{X_1}+\tilde{F}_{X_1}, \nonumber\\
&\geq (1+\tilde{\Phi}) C_{X-X_1} +  C_{X_1}+\tilde{\Phi} C_{X_1}, \text{by $\tilde{F}_{S} \geq \tilde{\Phi} C_{S}$,}\nonumber\\
&=(1+\tilde{\Phi}) C_{X}.
\nonumber
\end{align}

Since there is a cut of size $(1+\tilde{\Phi})C_{V^+}$, we have proved that the min-cut is equal to $(1+\tilde{\Phi}) C_{V^+}$. By the max-flow min-cut theorem \cite{Alon}, there exists a flow function $h_{ij}$ for all directed edges in the network so that $h_{ij}$ is bounded above by the capacity of $(i,j)$. Also, for each fixed $x \in X$ and $y \in Y$, we have
\begin{align}
\sum_{j \in Y} h_{xj} & =(1+\tilde{\Phi}) \pi_x, \\
\sum_{j \in X} h_{jy}& \leq \pi_y.
\end{align}
Recall \eqref{lowerbound}
\[ \lambda_1 - \lambda_0 \geq \frac{\sum_{i<j} w_{ij}(\hat{e}_i-\hat{e}_j)^2}{\sum_{i\in V^+} \pi_i \hat{e}^2_i}, \]
and note that 
\[
\lambda_1 - \lambda_0 \geq \frac{\sum_{i<j} w_{ij}(\hat{e}_i-\hat{e}_j)^2}{\sum_{i\in V^+} \pi_i \hat{e}^2_i} \times \frac{\sum_{\substack{i<j \\ \{i,j\}\in \tilde E}} \frac{h^2_{ij}}{w_{ij}}(\hat{e}_i+\hat{e}_j)^2}{\sum_{\substack{i<j\\ \{i, j\}\in \tilde E}} w_{ij} (\hat{e}_i+\hat{e}_j)^2}.
\]
At this point we have introduced the reduced graph $\tilde G$ through the summation over the edge set $\tilde E$. This has been possible as we are allowed to insert any number in the right hand side of the above inequality as long as it is smaller or equal to one. Exactly this mathematical freedom allowed us to generalise the Cheeger inequality.

In the denominator
  \begin{align}
 {\sum_{\substack{i<j \\ \{i,j\}\in \tilde E}} w_{ij} (\hat{e}_i+\hat{e}_j)^2} &\leq {2 \sum_{\substack{ \{i,j\}\in \tilde E}} w_{ij} (\hat{e}^2_i+\hat{e}^2_j)} \\
&= { \sum_{(i,j)\in \tilde E } w_{ij} (\hat{e}^2_i+\hat{e}^2_j)}\\
&=  2{\sum_{i \in V} \hat{e}^2_i \sum_{\substack{j\\(i,j)\in \tilde E }} w_{ij} }\\
&=2{\sum_{i \in V} \pi_i \hat{e}^2_i {c_i}}\\
&\leq 2c \sum_{i \in V^+} \pi_i \hat{e}^2_i. 
 \label{bottom1}
  \end{align}

In the numerator
\begin{align}
&{\sum_{\substack{i< j\\E}} w_{ij}(\hat{e}_i-\hat{e}_j)^2}{\sum_{\substack{i< j\\\tilde E}} \frac{h^2_{ij}}{w_{ij}}(\hat{e}_i+\hat{e}_j)^2} \\ 
&\geq {\sum_{\substack{i<j \\ \tilde E}} w_{ij}(\hat{e}_i-\hat{e}_j)^2}{\sum_{\substack{i<j\\ \tilde E}} \frac{h^2_{ij}}{w_{ij}}(\hat{e}_i+\hat{e}_j)^2} \\
&\geq (\sum_{\substack{i< j \\ \tilde E}} h_{ij} (\hat{e}_i^2-\hat{e}_j^2))^2, \text{by Cauchy-Schwarz}; \\
&= (\sum_{\substack{i\leq j\\ \tilde E}} h_{ij} (\hat{e}_i^2-\hat{e}_j^2))^2, \\
&= (\sum_{\{i,j\}\in \tilde E} h_{ij} (\hat{e}_i^2-\hat{e}_j^2))^2, \text{by using $\sum_{i\leq j}=\sum_{\{i,j\}}$}; \\
&= (\sum_{i\in V} \hat{e}_i^2 (\sum_{\substack{j \\ \{i,j\} \in \tilde E}} h_{ij} - \sum_{{\substack{j \\ \{i,j\} \in \tilde E}} }h_{ji}))^2\\
&\geq (\sum_{i\in V^+} \hat{e}_i^2 ((1+\tilde{\Phi}) \pi_i - \pi_i))^2\\
&\geq (\sum_{i\in V^+} \hat{e}_i^2 \pi_i)^2 \tilde{\Phi}^2.
  \label{top1}
\end{align}

Putting \eqref{top1} and \eqref{bottom1} together we get
 \[  \lambda_1 - \lambda_0 \geq \frac{\tilde{\Phi}^2}{2 c}.\]
 as required. \;\; ${\blacksquare}$


\begin{thebibliography}{11}

\bibitem{Nielsen}
H. Haselgrove, M. A. Nielsen, T. J. Osborne, Entanglement, correlations, and the energy gap in many-body quantum systems, Phys. Rev. A $\bf{69}$, 032303 (2004)

\bibitem{Wim}
D. Aharonov, Wim van Dam, J. Kempe, Z. Landau, S. Lloyd, O. Regev, Adiabatic Quantum Computation is Equivalent to Standard Quantum Computation, SIAM Journal of Computing, {\bf 37}, 166-194 (2007).

\bibitem{Freedman} 
M. Freedman, C. Nayak and K. Shtengel, Lieb-Schultz-Mattis theorem for quasi-topological systems, Phys. Rev. B {\bf 78}, 174411 (2008).

\bibitem{Cheeger}
{J. Cheeger}, {A lower bound for the smallest eigenvalue of the Laplacian}. Problems in analysis, Princeton Univ. Press, Princeton, N. J. (1970).

\bibitem{Chung} 
F. R. K. Chung, {\em Spectral Graph Theory}, CBMS, AMS (1997).

\bibitem{Schuch} 
N. Schuch, D. Perez-Garcia, and I. Cirac,
 arXiv:1010.3732 (2010).

\bibitem{Alon}{N. Alon,} \text{Eigenvalues and Expanders,} COMBINATORICA, {\bf 6}, 83-96 (1986).

\bibitem{Golovach}
P. A. Golovach, Computing the isoperimetric number of a graph, Cypernetics and systems analysis,
{\bf 30}, 453-457 (1994).

\bibitem{Horn}
R. Horn and C. Johnson, {\em Matrix Analysis,} Cambridge Univ. Press, Cambridge, (1985).

\bibitem{Sinclair}
{A. Sinclair and M. Jerrum}, {Approximate Counting, Uniform Generation and Rapidly Mixing Markov Chains}, volume 314, {\em Lecture Notes in Comput. Sci.,} Springer, Berlin, (1988).

\bibitem{Abbas} A. Al-Shimary and  J.K. Pachos, in preparation.

\end{thebibliography}
\end{document}